\documentclass[
twocolumn,
english,
superscriptaddress,
amsmath,
floatfix,
longbibliography,
floats,
noeprint,
prl,
aps,
10pt,
]{revtex4-2}

\usepackage[T1]{fontenc}
\usepackage{lmodern}     
\usepackage{microtype}   
\usepackage{bm}

\usepackage{mathtools}   
\usepackage{amssymb}
\usepackage{braket}
\usepackage{mathrsfs}    

\usepackage{graphicx}
\usepackage[table]{xcolor}
\usepackage{array}
\usepackage{booktabs}     
\usepackage{multirow}
\usepackage{diagbox}
\usepackage{dcolumn}      
\usepackage{makecell}
\usepackage{siunitx}

\AddToHook{env/table/begin}{\setlength{\belowcaptionskip}{6pt}}
\usepackage{indentfirst}
\usepackage[normalem]{ulem} 
\usepackage{ragged2e}
\usepackage{enumitem}     

\usepackage{tikz}
\usetikzlibrary{quantikz2,trees,arrows,positioning,automata,shadows,fit,shapes,cd,decorations.markings}

\usepackage{xurl}
\usepackage[colorlinks=true,linkcolor=magenta,citecolor=magenta,urlcolor=magenta]{hyperref}
\usepackage{etoolbox}
\AtBeginEnvironment{thebibliography}{%
\renewcommand{\bibinfo}[2]{\ifstrequal{#1}{title}{\textcolor{black}{#2}}{#2}}%
}

\newcommand{\OU}{Graduate School of Engineering Science, The University of Osaka, 1-3 Machikaneyama, Toyonaka, Osaka 560-8531, Japan}
\newcommand{\QIQB}{Center for Quantum Information and Quantum Biology, The University of Osaka, Toyonaka, 560-0043, Japan}
\newcommand{\RQC}{Center for Quantum Computing, RIKEN, Wako Saitama 351-0198, Japan}

\newcommand{\KU}{Graduate School of Informatics, Kyoto University, Sakyo-ku, Kyoto, 606-8501, Japan}

\newcommand{\SWAP}{\mathrm{SWAP}}

\newcommand{\Stot}{\mathbf{S}_{\mathrm{tot}}}

\newcommand{\SUtwo}{\mathrm{SU(2)}}

\begin{document}

\title{Low-Depth Initial-State Preparation for Ground-State Energy Estimation of Two-Dimensional Strongly Correlated Systems}

\author{Ryo Watanabe}
\affiliation{\OU}

\author{Keisuke Fujii}
\affiliation{\OU}
\affiliation{\KU}
\affiliation{\QIQB}
\affiliation{\RQC}

\begin{abstract}
Quantum algorithms for ground-state energy estimation require initial states with non-negligible fidelity to the ground state.
Guided by classical simulations, we design shallow circuits for the two-dimensional half-filled Hubbard model by first preparing an approximate Heisenberg ground state and then applying charge-fluctuation gates derived directly from the Schrieffer--Wolff (SW) generator.
For the Heisenberg model, we demonstrate effective parameter transfer from a $4\times4$ lattice to lattices up to $10\times10$ without further optimization.
We obtain promising lower bounds on the ground-state fidelity using tensor-network simulations, variational Monte Carlo reference states, and estimates of the ground-state energy and singlet gap.
Exact $4\times4$ calculations show that the SW gates substantially improve the Hubbard ground-state fidelity of the embedded Heisenberg state.
The successful Heisenberg parameter transfer supports the use of these small-lattice results to design shallow Hubbard circuits on larger lattices beyond the reach of classical simulations.
For a $10\times10$ lattice, the estimated preparation cost is on the order of $10^5$ $T$ gates, including Clifford+$T$ synthesis, which is well within the megaquop regime.
These results offer a route to low-cost initial-state preparation for ground-state energy estimation in strongly correlated systems.
\end{abstract}

\maketitle

\section{Introduction}

Quantum computers offer a way to address computational problems believed to be intractable classically~\cite{Feynman1982,Lloyd1996,Shor1997_prime_factorization}.
One such application is ground-state energy estimation for quantum many-body systems, a central challenge in condensed matter physics and quantum chemistry.
Various algorithms have been developed for ground-state energy estimation on fault-tolerant quantum computers (FTQCs)~\cite{PhysRevLett.77.3260,PhysRevA.57.127}, including quantum phase estimation (QPE)~\cite{kitaev1995quantummeasurementsabelianstabilizer,PhysRevLett.83.5162}, its variants~\cite{10.1098/rspa.1998.0164,PhysRevA.75.012328}, and spectral filtering methods~\cite{ge2018fastergroundstatepreparation,Lin2020nearoptimalground}.
Progress toward FTQC has also drawn attention to the early-FTQC regime, where logical qubits and magic states are limited resources~\cite{PRXQuantum.5.020101}.
This has driven the development of resource-efficient algorithms that reduce ancilla requirements, shorten coherent circuit depth, and shift part of the energy extraction to classical post-processing~\cite{Somma_2019,PRXQuantum.3.010318,Zhang2022computingground,PRXQuantum.4.020331}.

These algorithms typically require an initial state that has a non-negligible fidelity to the ground state, since ground-state energy estimation is QMA-hard in general~\cite{Kempe2006_QMA,Aharonov2009,QMA_complexity}.
While this fidelity affects the overall runtime of quantum algorithms~\cite{Lemieux2021Resources,Yoshioka2024}, theoretical analyses often treat state preparation as a black box without accounting for its explicit cost~\cite{PhysRevX.15.021057,kanasugi2026enablingchemicallyaccuratequantum}.
In QPE, for example, the probability of obtaining the ground-state energy is given by the fidelity $F_0$ of the initial state with the ground state.
Consequently, an average of $1/F_0$ repetitions is required, while $F_0$ can decay exponentially with system size.
State-preparation circuits must therefore balance circuit complexity against target fidelity to optimize end-to-end performance, which is especially critical in the early-FTQC regime.

One approach is to compute an accurate classical approximation of the ground state and compile it into a quantum circuit.
This strategy has been investigated using sums of Slater determinants~\cite{tubman2018postponingorthogonalitycatastropheefficient} and matrix product states (MPS) optimized by the density-matrix renormalization group method~\cite{PhysRevLett.132.040404,PRXQuantum.5.040339,PRXQuantum.6.020327}.
Such states are typically mapped to circuits using linear combinations of unitaries~\cite{Gilyen2019_QSVT} or unitary synthesis techniques~\cite{Low2024tradingtgatesdirty}.
However, achieving the desired accuracy can require circuits that are too deep for near-term hardware or devices with limited resources, as a compact classical representation does not necessarily translate into a short quantum circuit.

Parameterized quantum circuits (PQCs) provide compact ansatzes that can be tailored to prepare initial states for ground-state energy estimation algorithms~\cite{PhysRevA.92.042303,Ho_2019,Cade_2020}.
The variational quantum eigensolver (VQE) established PQCs as a practical approach to estimating ground-state energies.
In this hybrid quantum--classical algorithm, a quantum computer evaluates variational energies and a classical optimizer updates the circuit parameters to minimize the energy~\cite{Peruzzo2014,Cerezo2021,Tilly_2022}.

VQE has garnered considerable attention as a leading application for noisy intermediate-scale quantum devices~\cite{Preskill2018quantumcomputingin}.
However, the high cost of energy and gradient evaluations remains a substantial hurdle~\cite{PhysRevA.98.032309,Schuld_2019}, and quantum advantage has yet to be demonstrated.
Furthermore, many circuits within reach of current hardware can be efficiently simulated classically to a good approximation~\cite{Valiant2002Matchgates,TerhalDiVincenzo2002,Vidal2003EfficientSimulation,BravyiGosset2016,Bravyi2019simulationofquantum,Rall2019PauliPropagation,Napp2022ShallowCircuits,Angrisani2025NoiselessCircuits}, particularly with tensor-network methods~\cite{MarkovShi2008,Zhou2020Simulation,GrayKourtis2021,Ayral2023DMRGCircuits,GrayChan2024}.
In such regimes, classical simulations can provide the energy and gradient evaluations needed for VQE, thereby reducing or even replacing quantum resource usage~\cite{FrancaGarciaPatron2021,Tindall2024Eagle,Patra2024IBM,Begusic2024Fast,Fontana2025NoisyVQC,Gustafson2025Surrogate,watanabe2026tensornetworksurrogatemodels}.

In this work, we draw on classical simulation techniques to design shallow quantum circuits for ground-state preparation in the two-dimensional Heisenberg and half-filled Hubbard models.
To make these circuits hardware-friendly, we leverage native device connectivity and gate sets to prioritize circuit-depth reduction, rather than restricting the ansatz to classically tractable forms~\cite{gibbs2025learningcircuitsinfinitetensor,gibbs2026lowtcountpreparationnuclear}.
Specifically, we exploit the strong-coupling connection between the two models via the Schrieffer--Wolff (SW) transformation~\cite{Bravyi2011SW}, constructing the Hubbard circuit by dressing an approximate Heisenberg ground-state preparation circuit with charge-fluctuation gates.
While Murta and Fern\'andez-Rossier employed Baeriswyl-type charge-fluctuation operators~\cite{PhysRevB.109.035128}, we derive these gates directly from the SW generator, treating their rotation angles as variational parameters.
This strategy extends the Heisenberg circuit to the Hubbard model with only a minimal depth overhead.

For the Heisenberg model, we evaluate parameter transferability in terms of ground-state fidelity.
Parameters optimized on a $4\times4$ lattice yield fidelity lower bounds that remain robust for lattices up to $10\times10$ without further optimization.
To obtain these bounds, we combine two-dimensional tensor-network simulations~\cite{VerstraeteMurgCirac2008,Cirac2021MPSPEPS} with variational Monte Carlo (VMC) states~\cite{Foulkes2001QMC,BeccaSorella2017QMC,Misawa2019mVMC,Szoldra2023FidelityOverlap}.
Specifically, we quantify the proximity of the VMC reference to the ground state via its energy, alongside estimates of the ground-state energy and singlet gap. The overlap between this reference and the circuit state then provides a lower bound on the circuit's ground-state fidelity.
This enables benchmarking circuits even when direct comparison with the exact ground state is intractable.
Across the regimes studied, these bounds systematically improve with circuit depth, motivating further exploration of parameter transfer up to depths where small-system performance saturates~\cite{watanabe2026tensornetworksurrogatemodels}.

For the Hubbard model, exact $4\times4$ calculations show that incorporating SW charge-fluctuation gates enhances the ground-state fidelity, with further gains achieved by optimizing their angles.
Combined with the Heisenberg transfer results, these findings motivate extending this construction to regimes beyond the reach of efficient classical simulation.
For both models, the $10\times10$ circuits examined here incur preparation costs on the order of $10^5$ $T$ gates, assuming Clifford+$T$ synthesis at a per-rotation error of $10^{-5}$.
These costs fall well below those reported for subsequent ground-state energy estimation~\cite{Yoshioka2024,Kivlichan2020improvedfault,Campbell2022Hubbard,Pathak2023,apel2026compiling2dfermihubbardgroundstate}, positioning these circuits as attractive candidates for initial-state preparation in the megaquop regime of early-FTQC~\cite{chung2026partiallyfaulttolerantquantumcomputation}.

The remainder of this paper is organized as follows.
Section~\ref{sec:variational_quantum_circuits} introduces the circuit construction.
Section~\ref{sec:direct_fidelity_estimation} derives the fidelity estimator with respect to the VMC reference state and relates it to ground-state fidelity bounds via the singlet gap.
Section~\ref{sec:results} details the numerical implementation and reports the Heisenberg parameter-transfer results, the Hubbard circuit benchmarks, and the circuit costs.
Section~\ref{sec:conclusion} concludes the paper.
In addition, Appendix~\ref{sec:direct_fidelity_optimization} provides a demonstration of gradient-based optimization using our tensor network simulations for the Heisenberg circuit.

\section{Parameterized Quantum Circuits}

\label{sec:variational_quantum_circuits}

We consider an $L_x\times L_y$ square lattice with open boundary conditions (OBC), where $L_x$ and $L_y$ are even.
The Hubbard Hamiltonian is given by
\begin{equation}
\label{eq:Fermi--Hubbard_Hamiltonian}
H
=-t\sum_{\braket{i,j}\in\mathcal{B},\,\sigma}
\left(c^\dagger_{i,\sigma}c_{j,\sigma}
+c^\dagger_{j,\sigma}c_{i,\sigma}\right)
+U\sum_{i\in\Lambda}n_{i,\uparrow}n_{i,\downarrow},
\end{equation}
with $t>0$ and $U/t=8$.
At this coupling, the interaction strength equals the thermodynamic-limit bandwidth $W=8t$ of the square lattice~\cite{Hubbard1963ElectronCorrelations}.
Here, $\Lambda$ denotes the set of lattice sites with cardinality $N \equiv |\Lambda|=L_xL_y$, and $\mathcal{B}$ denotes the set of nearest-neighbor bonds $\braket{i,j}$ ($i,j\in\Lambda$), each counted once.
The operators $c^\dagger_{i,\sigma}$ and $c_{i,\sigma}$ are the standard fermionic creation and annihilation operators for an electron with spin $\sigma\in\{\uparrow,\downarrow\}$ at site $i$, and $n_{i,\sigma}=c^\dagger_{i,\sigma}c_{i,\sigma}$ is the corresponding number operator.
At half filling, the number of electrons equals the number of lattice sites $N$.

In the strong-coupling limit ($U/t \gg 1$), charge fluctuations are strongly suppressed, restricting the low-energy physics predominantly to the singly occupied Hilbert space.
Applying the SW transformation to second order in $t/U$ (Appendix~\ref{sec:SW_transformation}) gives, up to an additive constant, the antiferromagnetic spin-$\tfrac{1}{2}$ Heisenberg Hamiltonian
\begin{equation}
\widetilde H
=
J \sum_{\braket{i,j} \in \mathcal{B}}
\mathbf{S}_i \cdot \mathbf{S}_j~,
\label{eq:Heisenberg_model}
\end{equation}
with an exchange coupling $J = 4t^2/U$~\cite{Bravyi2011SW}.
Here, the spin operators are given by $\mathbf{S}_i = (S^x_i, S^y_i, S^z_i) = \tfrac{1}{2} c_i^\dagger \boldsymbol{\sigma} c_i$, where $c_i = (c_{i,\uparrow}, c_{i,\downarrow})^{\mathsf T}$ denotes the spinor of fermionic annihilation operators and $\boldsymbol{\sigma} = (\sigma_x, \sigma_y, \sigma_z)$ is the vector of Pauli matrices.

Our goal is to design quantum circuits capable of accurately approximating the ground state even at shallow circuit depths. To achieve this, we leverage the Heisenberg model derived from the strong-coupling Hubbard model to construct physically motivated PQCs that reflect the two-dimensional lattice connectivity while preserving the model's symmetries. In what follows, we first outline the variational ansatz and its physical rationale, and then detail its implementation as a quantum circuit.

\subsection{Variational ansatz}\label{sec:variational_ansatz}

The Heisenberg Hamiltonian $\widetilde H$ exhibits $\SUtwo$ symmetry, commuting with all components of the total spin operator:
\begin{equation}
[\widetilde H, \Stot^{\alpha}] = 0 \quad (\alpha \in \{x, y, z\}),
\end{equation}
where $\Stot^{\alpha} = \sum_{i \in \Lambda} S^{\alpha}_i$.
Consequently, $\widetilde H$ also commutes with the total spin squared, $\Stot^2 = \left(\sum_{i \in \Lambda} \mathbf{S}_i\right)^2$.
The eigenstates can therefore be labeled by the total spin quantum number $S$,
\begin{equation}
\Stot^2 \ket{\psi} = S(S+1) \ket{\psi}.
\label{eq:eigen_Stot}
\end{equation}

The Lieb--Mattis theorem states that the ground state of the spin-$\tfrac{1}{2}$ antiferromagnetic Heisenberg model on a bipartite lattice $\Lambda = \Lambda_{\mathrm{A}} \sqcup \Lambda_{\mathrm{B}}$ has a total spin of $S = \tfrac{1}{2}\bigl| |\Lambda_{\mathrm{A}}| - |\Lambda_{\mathrm{B}}| \bigr|$~\cite{10.1063/1.1724276}.
For the square lattice considered here, with even $L_x$ and $L_y$, the two sublattices have the same size and hence $S = 0$.

We encode each spin onto one qubit, with $\ket{\uparrow}=\ket0$ and $\ket{\downarrow}=\ket1$, and take a product of singlet dimers as the initial state:
\begin{equation}
\ket{\Phi}
=
\bigotimes_{\braket{i,j}\in\mathcal{B}_0}
\ket{\mathrm{s}_{ij}}~,
\label{eq:dimer_products}
\end{equation}
where $\ket{\mathrm{s}_{ij}} \equiv \frac{1}{\sqrt{2}}\left(\ket{\uparrow_i\downarrow_j}-\ket{\downarrow_i\uparrow_j}\right)$ denotes a spin-singlet pair, and $\mathcal{B}_0\subseteq\mathcal{B}$ is a perfect matching on the lattice $\Lambda$, that is, a set of disjoint bonds covering every site of $\Lambda$ exactly once.
$\mathcal{B}_0$ pairs every spin into a singlet by construction, placing the resulting state $\ket{\Phi}$ in the $S=0$ subspace.

\begin{figure}[t]
\centering
\includegraphics[clip,width=0.9\linewidth]{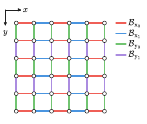}
\caption{
A bond partition of a $6 \times 6$ open-boundary square lattice into four matchings, $\mathcal{B}_{\mathrm{x}_0}$, $\mathcal{B}_{\mathrm{x}_1}$, $\mathcal{B}_{\mathrm{y}_0}$, and $\mathcal{B}_{\mathrm{y}_1}$ (colors indicated in the legend).
For general lattice sizes with even $L_x$ and $L_y$, $\mathcal{B}_{\mathrm{x}_0}$ and $\mathcal{B}_{\mathrm{y}_0}$ form perfect matchings covering all sites, while $\mathcal{B}_{\mathrm{x}_1}$ and $\mathcal{B}_{\mathrm{y}_1}$ leave boundary sites uncovered.
}
\label{fig:perfect_matching}
\end{figure}

As illustrated in Fig.~\ref{fig:perfect_matching}, the bond set is partitioned into four subsets.
If we number the columns from left to right and the rows from top to bottom starting from zero, each horizontal bond is assigned to $\mathcal{B}_{\mathrm{x}_0}$ or $\mathcal{B}_{\mathrm{x}_1}$ according to whether its left site lies in an even or odd column.
Likewise, each vertical bond is assigned to $\mathcal{B}_{\mathrm{y}_0}$ or $\mathcal{B}_{\mathrm{y}_1}$ according to whether its top site lies in an even or odd row, yielding the disjoint union $\mathcal{B} = \mathcal{B}_{\mathrm{x}_0} \sqcup \mathcal{B}_{\mathrm{x}_1} \sqcup \mathcal{B}_{\mathrm{y}_0} \sqcup \mathcal{B}_{\mathrm{y}_1}$.
Since $L_x$ and $L_y$ are even, the horizontal subset $\mathcal{B}_{\mathrm{x}_0}$ (red) and the vertical subset $\mathcal{B}_{\mathrm{y}_0}$ (green) each form a perfect matching.

We evolve the state $\ket{\Phi}$ via a sequence of exponentiated SWAP (eSWAP) gates, defined as
\begin{equation}
\mathcal{U}^{\braket{k,\ell}}(\theta) = \exp\left(-i\frac{\theta}{2}\SWAP_{\braket{k,\ell}}\right),
\label{eq:eSWAP}
\end{equation}
with a tunable angle $\theta \in \mathbb{R}$, where $\SWAP_{\braket{k,\ell}}=\frac{1}{2}I+2\mathbf{S}_k \cdot \mathbf{S}_\ell$.
Because the eSWAP gate respects $\SUtwo$ symmetry, the state remains strictly within the total spin-singlet ($S=0$) subspace throughout the evolution.
Physically, successively applying these gates delocalizes the initial product state into a superposition of singlets spanning the entire lattice~\cite{PhysRevA.101.052340}.
This procedure prepares an approximate resonating-valence-bond (RVB) state, a hallmark of the antiferromagnetic Heisenberg ground state on finite bipartite square lattices~\cite{PhysRevLett.61.365}.

The eSWAP gates for each matching form a gate layer.
We apply these layers in the following sequence:
\begin{equation}
\mathcal{B}_{\mathrm{y}_1}
\rightarrow
\mathcal{B}_{\mathrm{y}_0}
\rightarrow
\mathcal{B}_{\mathrm{x}_1}
\rightarrow
\mathcal{B}_{\mathrm{x}_0}~.
\label{eq:order}
\end{equation}
We apply the $\mathcal{B}_{\mathrm{x}_0}$ layer last to ensure it acts after the preceding layers have generated a superposition of distinct singlet configurations.
Since the initial state $\ket{\Phi}$ is a product of singlets defined on $\mathcal{B}_{\mathrm{x}_0}$, applying the $\mathcal{B}_{\mathrm{x}_0}$ layer directly to $\ket{\Phi}$ would introduce only a global phase.

The unitary operator for each layer is given by
\begin{equation}
\mathcal U^{\mathcal B_\nu}(\bm\theta_\nu)
=\prod_{\braket{i,j}\in\mathcal B_\nu}\mathcal U^{\braket{i,j}}(\theta_{\braket{i,j}}),
\label{eq:layer_swap}
\end{equation}
where $\nu\in\{\mathrm x_0,\mathrm x_1,\mathrm y_0,\mathrm y_1\}$ and $\bm\theta_\nu$ denotes the set of angles for that matching.
Because the bonds are disjoint, the gates within each layer can be executed in parallel.
These four layers are assembled into a single circuit block in the order specified by Eq.~\eqref{eq:order}:
\begin{equation}
\begin{split}
\mathcal{U}^{\mathcal{B}}(\bm{\theta}_p)
=
&\,\mathcal{U}^{\mathcal{B}_{\mathrm{x}_0}}(\bm{\theta}_{p,4})\,
\mathcal{U}^{\mathcal{B}_{\mathrm{x}_1}}(\bm{\theta}_{p,3})
\\
&\times
\mathcal{U}^{\mathcal{B}_{\mathrm{y}_0}}(\bm{\theta}_{p,2})\,
\mathcal{U}^{\mathcal{B}_{\mathrm{y}_1}}(\bm{\theta}_{p,1}).
\end{split}
\label{eq:block}
\end{equation}
Here, $p$ labels the block and the layer index $\ell=1,\ldots,4$ follows Eq.~\eqref{eq:order}.
Applying this block $N_{\mathrm b}$ times yields the Heisenberg circuit state
\begin{equation}
\ket{\Psi(\bm\theta)}=
\mathcal U^{\mathcal B}(\bm\theta_{N_{\mathrm b}})\cdots
\mathcal U^{\mathcal B}(\bm\theta_1)\ket\Phi,
\label{eq:variational_state}
\end{equation}
where $\bm\theta_p$ comprises the angles in block $p$, and $\bm\theta=(\bm\theta_1,\ldots,\bm\theta_{N_{\mathrm b}})$ denotes the full set of variational parameters.

To represent the resulting spin states in terms of fermionic degrees of freedom, we map each spin state $\ket{\varphi}=a\ket{\uparrow}+b\ket{\downarrow}$ onto the singly occupied subspace of the local occupation-number basis $\ket{n_{i,\uparrow}n_{i,\downarrow}}$~\cite{PhysRevB.109.035128}.
This is achieved via a local unitary $\mathcal{V}$ acting on the physical spin qubit and an ancillary qubit initialized to $\ket{0}$:
\begin{equation}
\mathcal{V}\bigl(\ket{0}\otimes\ket{\varphi}\bigr)
=
a\ket{10}+b\ket{01}~.
\label{eq:V_mapping}
\end{equation}
Here, $\ket{\uparrow}$ and $\ket{\downarrow}$ correspond to $\ket{10}$ and $\ket{01}$, respectively; the empty and doubly occupied states are $\ket{00}$ and $\ket{11}$.
The overall circuit thus requires $2N$ qubits: $N$ qubits encoding the spins and $N$ ancillae initialized to $\ket{0}$.
We denote a many-body configuration by $\bm{s}=(s_i)_{i\in\Lambda}$, with $s_i$ specifying the local spin or occupation state.
Under a site-by-site ordering of the Fock basis $\ket{\bm{s}} = \bigotimes_{i\in\Lambda} \ket{n_{i,\uparrow}n_{i,\downarrow}}$, $\mathcal{V}$ operates strictly locally on each site.
Consequently, it requires no reordering of fermionic operators between sites, thereby incurring no additional sign factors.

At $U/t = 8$, the Hubbard ground state contains configurations with doubly occupied sites (doublons) and empty sites (holons) that cannot be captured by a purely spin-based description within an effective model.
To incorporate these configurations, we augment the ansatz with charge-fluctuation gates.
As derived in Appendix~\ref{sec:SW_transformation}, the first-order SW transformation is generated by the anti-Hermitian operator
\begin{equation}
\Omega
=
-\frac{t}{U}
\sum_{\braket{i,j}\in\mathcal{B},\sigma}
\bigl(n_{i,\bar\sigma}-n_{j,\bar\sigma}\bigr)
\bigl(
c^\dagger_{i,\sigma} c_{j,\sigma}
-
c^\dagger_{j,\sigma} c_{i,\sigma}
\bigr)~,
\label{eq:SW_generator_explicit}
\end{equation}
where $\bar\sigma$ denotes the spin opposite to $\sigma$.
Applying $e^{-\Omega}$ to the Heisenberg ground state, embedded via Eq.~\eqref{eq:V_mapping}, reproduces the Hubbard ground state up to first order in $t/U$.

Using a first-order Suzuki--Trotter decomposition~\cite{Trotter_1959,Suzuki1976}, we approximate $e^{-\Omega}$ by a product of two-site unitaries defined on the four matchings (Fig.~\ref{fig:perfect_matching}).
Each local unitary takes the form
\begin{equation}
\mathcal{W}^{\braket{i,j}}(\phi) = \exp \left(-i \frac{\phi}{2} \omega^{\braket{i,j}}\right)~,
\label{eq:charge_gate}
\end{equation}
with
\begin{equation}
\omega^{\braket{i,j}}
=
\frac{i}{2}
\sum_{\sigma}
\bigl(n_{i,\bar\sigma}-n_{j,\bar\sigma}\bigr)
\bigl(
c^\dagger_{i,\sigma} c_{j,\sigma}
-
c^\dagger_{j,\sigma} c_{i,\sigma}
\bigr)~.
\label{eq:omega_local}
\end{equation}
Setting $\phi=4t/U$ reproduces the first-order SW transformation up to a Trotter error of $\mathcal{O}((t/U)^2)$.

The Hubbard Hamiltonian $H$ also possesses $\SUtwo$ symmetry and conserves the total particle number $\sum_{i,\sigma} n_{i,\sigma}$ with eigenvalue $N_{\mathrm{e}}$.
At half filling, Lieb's theorem ensures that the total spin satisfies $S = \tfrac{1}{2}\bigl| |\Lambda_{\mathrm{A}}| - |\Lambda_{\mathrm{B}}| \bigr|$, consistent with the Lieb--Mattis theorem for $\widetilde H$~\cite{PhysRevLett.62.1201}; thus, the $S=0$ ground-state property of the Heisenberg model persists even at finite $U$.
The embedding $\mathcal{V}$ in Eq.~\eqref{eq:V_mapping} initializes each site with a single fermion, while the charge gates preserve both the particle number and total spin.
Consequently, the variational search is confined to the $(N_{\mathrm{e}},S)=(N,0)$ subspace containing the target ground state.

The charge-fluctuation gates form layers on the same matchings:
\begin{equation}
\mathcal W^{\mathcal B_\nu}(\bm\phi_\nu)
=\prod_{\braket{i,j}\in\mathcal B_\nu}\mathcal W^{\braket{i,j}}(\phi_{\braket{i,j}}),
\label{eq:layer_sw}
\end{equation}
where $\bm\phi_\nu$ denotes the charge-gate angles on $\mathcal B_\nu$.
Applying these four layers in the order specified by Eq.~\eqref{eq:order} gives the charge-fluctuation block:
\begin{equation}
\begin{split}
\mathcal{W}^{\mathcal{B}}(\bm{\phi})
=
&\,\mathcal{W}^{\mathcal{B}_{\mathrm{x}_0}}(\bm{\phi}_4)\,
\mathcal{W}^{\mathcal{B}_{\mathrm{x}_1}}(\bm{\phi}_3)
\\
&\times
\mathcal{W}^{\mathcal{B}_{\mathrm{y}_0}}(\bm{\phi}_2)\,
\mathcal{W}^{\mathcal{B}_{\mathrm{y}_1}}(\bm{\phi}_1)~.
\end{split}
\label{eq:charge_block}
\end{equation}
This leads to the variational ansatz
\begin{equation}
\begin{split}
\ket{\Psi(\bm{\theta},\bm{\phi})}
=
&\,\mathcal{W}^{\mathcal{B}}(\bm{\phi})\,
\mathcal{V}^{\otimes N}
\Bigl[
\ket{0}^{\otimes N}
\otimes {}
\\
&\qquad
\mathcal{U}^{\mathcal{B}}(\bm{\theta}_{N_{\mathrm{b}}}) \cdots \mathcal{U}^{\mathcal{B}}(\bm{\theta}_1) \ket{\Phi}
\Bigr]~.
\end{split}
\label{eq:hubbard_variational_state}
\end{equation}
The vectors $\bm{\theta}_p$ and $\bm{\phi}$ comprise all angles associated with the $p$-th block and the charge-fluctuation layers, respectively.
Since each gate introduces one parameter, this setup yields $N_{\mathrm{b}}|\mathcal{B}|$ angles from the eSWAP blocks and $|\mathcal{B}|$ from the charge-fluctuation block, which collectively define the full parameter set
\begin{equation}
\bm{\mu} \equiv (\bm{\theta},\bm{\phi})~.
\label{eq:parameter_set}
\end{equation}

\subsection{Quantum circuit construction}\label{sec:quantum_circuit_construction}

The singlet-dimer product state $\ket{\Phi}$ in Eq.~\eqref{eq:dimer_products} is prepared by initializing the qubits to $\ket{0}^{\otimes N}$ and applying a Clifford circuit across each bond $\braket{i,j} \in \mathcal{B}_0$:
\begin{equation}
\begin{tikzpicture}[scale=1.000000,x=1pt,y=1pt]
\filldraw[color=white] (0.000000, -8.500000) rectangle (70.000000, 25.500000);
\draw[color=black] (0.000000,17.000000) -- (70.000000,17.000000);
\draw[color=black] (0.000000,0.000000) -- (70.000000,0.000000);
\begin{scope}
\draw[fill=white] (13.000000, 17.000000) +(-45.000000:9.899495pt and 9.899495pt) -- +(45.000000:9.899495pt and 9.899495pt) -- +(135.000000:9.899495pt and 9.899495pt) -- +(225.000000:9.899495pt and 9.899495pt) -- cycle;
\clip (13.000000, 17.000000) +(-45.000000:9.899495pt and 9.899495pt) -- +(45.000000:9.899495pt and 9.899495pt) -- +(135.000000:9.899495pt and 9.899495pt) -- +(225.000000:9.899495pt and 9.899495pt) -- cycle;
\draw (13.000000, 17.000000) node {$H$};
\end{scope}
\begin{scope}
\draw[fill=white] (13.000000, -0.000000) +(-45.000000:9.899495pt and 9.899495pt) -- +(45.000000:9.899495pt and 9.899495pt) -- +(135.000000:9.899495pt and 9.899495pt) -- +(225.000000:9.899495pt and 9.899495pt) -- cycle;
\clip (13.000000, -0.000000) +(-45.000000:9.899495pt and 9.899495pt) -- +(45.000000:9.899495pt and 9.899495pt) -- +(135.000000:9.899495pt and 9.899495pt) -- +(225.000000:9.899495pt and 9.899495pt) -- cycle;
\draw (13.000000, -0.000000) node {$X$};
\end{scope}
\draw (35.000000,17.000000) -- (35.000000,0.000000);
\begin{scope}
\draw[fill=white] (35.000000, 0.000000) circle(3.000000pt);
\clip (35.000000, 0.000000) circle(3.000000pt);
\draw (32.000000, 0.000000) -- (38.000000, 0.000000);
\draw (35.000000, -3.000000) -- (35.000000, 3.000000);
\end{scope}
\filldraw (35.000000, 17.000000) circle(1.500000pt);
\begin{scope}
\draw[fill=white] (57.000000, 17.000000) +(-45.000000:9.899495pt and 9.899495pt) -- +(45.000000:9.899495pt and 9.899495pt) -- +(135.000000:9.899495pt and 9.899495pt) -- +(225.000000:9.899495pt and 9.899495pt) -- cycle;
\clip (57.000000, 17.000000) +(-45.000000:9.899495pt and 9.899495pt) -- +(45.000000:9.899495pt and 9.899495pt) -- +(135.000000:9.899495pt and 9.899495pt) -- +(225.000000:9.899495pt and 9.899495pt) -- cycle;
\draw (57.000000, 17.000000) node {$Z$};
\end{scope}
\end{tikzpicture}

\label{eq:singlet_dimer_construction}
\end{equation}

Up to a global phase, the eSWAP gate in Eq.~\eqref{eq:eSWAP} is decomposed into a sequence of gates as
\begin{equation}
\begin{tikzpicture}[scale=1.000000,x=1pt,y=1pt]
\filldraw[color=white] (0.000000, -8.500000) rectangle (238.000000, 25.500000);
\draw[color=black] (0.000000,17.000000) -- (238.000000,17.000000);
\draw[color=black] (0.000000,0.000000) -- (238.000000,0.000000);
\draw (6.000000,17.000000) -- (6.000000,0.000000);
\begin{scope}
\draw[fill=white] (6.000000, 17.000000) circle(3.000000pt);
\clip (6.000000, 17.000000) circle(3.000000pt);
\draw (3.000000, 17.000000) -- (9.000000, 17.000000);
\draw (6.000000, 14.000000) -- (6.000000, 20.000000);
\end{scope}
\filldraw (6.000000, 0.000000) circle(1.500000pt);
\begin{scope}
\draw[fill=white] (22.000000, -0.000000) +(-45.000000:9.899495pt and 9.899495pt) -- +(45.000000:9.899495pt and 9.899495pt) -- +(135.000000:9.899495pt and 9.899495pt) -- +(225.000000:9.899495pt and 9.899495pt) -- cycle;
\clip (22.000000, -0.000000) +(-45.000000:9.899495pt and 9.899495pt) -- +(45.000000:9.899495pt and 9.899495pt) -- +(135.000000:9.899495pt and 9.899495pt) -- +(225.000000:9.899495pt and 9.899495pt) -- cycle;
\draw (22.000000, -0.000000) node {$S$};
\end{scope}
\begin{scope}
\draw[fill=white] (51.000000, -0.000000) +(-45.000000:22.627417pt and 9.899495pt) -- +(45.000000:22.627417pt and 9.899495pt) -- +(135.000000:22.627417pt and 9.899495pt) -- +(225.000000:22.627417pt and 9.899495pt) -- cycle;
\clip (51.000000, -0.000000) +(-45.000000:22.627417pt and 9.899495pt) -- +(45.000000:22.627417pt and 9.899495pt) -- +(135.000000:22.627417pt and 9.899495pt) -- +(225.000000:22.627417pt and 9.899495pt) -- cycle;
\draw (51.000000, -0.000000) node {$R_Y(\tfrac{\theta}{2})$};
\end{scope}
\draw (76.000000,17.000000) -- (76.000000,0.000000);
\begin{scope}
\draw[fill=white] (76.000000, 0.000000) circle(3.000000pt);
\clip (76.000000, 0.000000) circle(3.000000pt);
\draw (73.000000, 0.000000) -- (79.000000, 0.000000);
\draw (76.000000, -3.000000) -- (76.000000, 3.000000);
\end{scope}
\filldraw (76.000000, 17.000000) circle(1.500000pt);
\begin{scope}
\draw[fill=white] (105.000000, -0.000000) +(-45.000000:28.284271pt and 9.899495pt) -- +(45.000000:28.284271pt and 9.899495pt) -- +(135.000000:28.284271pt and 9.899495pt) -- +(225.000000:28.284271pt and 9.899495pt) -- cycle;
\clip (105.000000, -0.000000) +(-45.000000:28.284271pt and 9.899495pt) -- +(45.000000:28.284271pt and 9.899495pt) -- +(135.000000:28.284271pt and 9.899495pt) -- +(225.000000:28.284271pt and 9.899495pt) -- cycle;
\draw (105.000000, -0.000000) node {$R_Y(-\tfrac{\theta}{2})$};
\end{scope}
\draw (134.000000,17.000000) -- (134.000000,0.000000);
\begin{scope}
\draw[fill=white] (134.000000, 0.000000) circle(3.000000pt);
\clip (134.000000, 0.000000) circle(3.000000pt);
\draw (131.000000, 0.000000) -- (137.000000, 0.000000);
\draw (134.000000, -3.000000) -- (134.000000, 3.000000);
\end{scope}
\filldraw (134.000000, 17.000000) circle(1.500000pt);
\begin{scope}
\draw[fill=white] (150.000000, -0.000000) +(-45.000000:9.899495pt and 9.899495pt) -- +(45.000000:9.899495pt and 9.899495pt) -- +(135.000000:9.899495pt and 9.899495pt) -- +(225.000000:9.899495pt and 9.899495pt) -- cycle;
\clip (150.000000, -0.000000) +(-45.000000:9.899495pt and 9.899495pt) -- +(45.000000:9.899495pt and 9.899495pt) -- +(135.000000:9.899495pt and 9.899495pt) -- +(225.000000:9.899495pt and 9.899495pt) -- cycle;
\draw (150.000000, -0.000000) node {$S^{\dagger}$};
\end{scope}
\begin{scope}
\draw[fill=white] (150.000000, 17.000000) +(-45.000000:9.899495pt and 9.899495pt) -- +(45.000000:9.899495pt and 9.899495pt) -- +(135.000000:9.899495pt and 9.899495pt) -- +(225.000000:9.899495pt and 9.899495pt) -- cycle;
\clip (150.000000, 17.000000) +(-45.000000:9.899495pt and 9.899495pt) -- +(45.000000:9.899495pt and 9.899495pt) -- +(135.000000:9.899495pt and 9.899495pt) -- +(225.000000:9.899495pt and 9.899495pt) -- cycle;
\draw (150.000000, 17.000000) node {$X$};
\end{scope}
\begin{scope}
\draw[fill=white] (183.000000, 17.000000) +(-45.000000:28.284271pt and 9.899495pt) -- +(45.000000:28.284271pt and 9.899495pt) -- +(135.000000:28.284271pt and 9.899495pt) -- +(225.000000:28.284271pt and 9.899495pt) -- cycle;
\clip (183.000000, 17.000000) +(-45.000000:28.284271pt and 9.899495pt) -- +(45.000000:28.284271pt and 9.899495pt) -- +(135.000000:28.284271pt and 9.899495pt) -- +(225.000000:28.284271pt and 9.899495pt) -- cycle;
\draw (183.000000, 17.000000) node {$R_Z(-\tfrac{\theta}{2})$};
\end{scope}
\begin{scope}
\draw[fill=white] (216.000000, 17.000000) +(-45.000000:9.899495pt and 9.899495pt) -- +(45.000000:9.899495pt and 9.899495pt) -- +(135.000000:9.899495pt and 9.899495pt) -- +(225.000000:9.899495pt and 9.899495pt) -- cycle;
\clip (216.000000, 17.000000) +(-45.000000:9.899495pt and 9.899495pt) -- +(45.000000:9.899495pt and 9.899495pt) -- +(135.000000:9.899495pt and 9.899495pt) -- +(225.000000:9.899495pt and 9.899495pt) -- cycle;
\draw (216.000000, 17.000000) node {$X$};
\end{scope}
\draw (232.000000,17.000000) -- (232.000000,0.000000);
\begin{scope}
\draw[fill=white] (232.000000, 17.000000) circle(3.000000pt);
\clip (232.000000, 17.000000) circle(3.000000pt);
\draw (229.000000, 17.000000) -- (235.000000, 17.000000);
\draw (232.000000, 14.000000) -- (232.000000, 20.000000);
\end{scope}
\filldraw (232.000000, 0.000000) circle(1.500000pt);
\end{tikzpicture}
,
\label{eq:eswap_decomposition}
\end{equation}
where $R_{P}(\alpha) = \exp(-i \frac{\alpha}{2} P)$ is the rotation gate for a Pauli operator $P$.

For the Hubbard circuit, the on-site embedding $\mathcal{V}$ in Eq.~\eqref{eq:V_mapping} is implemented using Clifford gates,
\begin{equation}
\vcenter{\hbox{\begin{tikzpicture}[scale=1.000000,x=1pt,y=1pt]
\filldraw[color=white] (0.000000, -8.500000) rectangle (52.000000, 25.500000);
\draw[color=black] (0.000000,17.000000) -- (52.000000,17.000000);
\draw[color=black] (0.000000,17.000000) node[left] {$\ket{0}$};
\draw[color=black] (0.000000,0.000000) -- (52.000000,0.000000);
\draw[color=black] (0.000000,0.000000) node[left] {$\ket{\varphi}$};
\begin{scope}
\draw[fill=white] (10.000000, -0.000000) +(-45.000000:9.899495pt and 9.899495pt) -- +(45.000000:9.899495pt and 9.899495pt) -- +(135.000000:9.899495pt and 9.899495pt) -- +(225.000000:9.899495pt and 9.899495pt) -- cycle;
\clip (10.000000, -0.000000) +(-45.000000:9.899495pt and 9.899495pt) -- +(45.000000:9.899495pt and 9.899495pt) -- +(135.000000:9.899495pt and 9.899495pt) -- +(225.000000:9.899495pt and 9.899495pt) -- cycle;
\draw (10.000000, -0.000000) node {$X$};
\end{scope}
\draw (26.000000,17.000000) -- (26.000000,0.000000);
\begin{scope}
\draw[fill=white] (26.000000, 17.000000) circle(3.000000pt);
\clip (26.000000, 17.000000) circle(3.000000pt);
\draw (23.000000, 17.000000) -- (29.000000, 17.000000);
\draw (26.000000, 14.000000) -- (26.000000, 20.000000);
\end{scope}
\filldraw (26.000000, 0.000000) circle(1.500000pt);
\begin{scope}
\draw[fill=white] (42.000000, -0.000000) +(-45.000000:9.899495pt and 9.899495pt) -- +(45.000000:9.899495pt and 9.899495pt) -- +(135.000000:9.899495pt and 9.899495pt) -- +(225.000000:9.899495pt and 9.899495pt) -- cycle;
\clip (42.000000, -0.000000) +(-45.000000:9.899495pt and 9.899495pt) -- +(45.000000:9.899495pt and 9.899495pt) -- +(135.000000:9.899495pt and 9.899495pt) -- +(225.000000:9.899495pt and 9.899495pt) -- cycle;
\draw (42.000000, -0.000000) node {$X$};
\end{scope}
\end{tikzpicture}
}}~.
\label{eq:V_implementation}
\end{equation}

Implementing the charge-fluctuation gate $\mathcal{W}^{\braket{i,j}}$ on quantum hardware requires mapping the fermionic creation and annihilation operators to Pauli operators.
Here, we derive an explicit representation using the Jordan--Wigner (JW) transformation~\cite{JordanWigner1928}.
We index the sites in the lattice $\Lambda$ as $i=0,\ldots,N-1$ in row-major order with $i=yL_x+x$.
Each spin-orbital is labeled by $(i,\sigma)$, where $i\in\Lambda$ and $\sigma\in\{\uparrow,\downarrow\}$, and is assigned a qubit index $\pi(i,\sigma) \in \{0,1,\ldots,2N-1\}$ according to the site-by-site ordering of the Fock basis:
\begin{equation}
\pi(i, \uparrow) = 2i, \quad \pi(i, \downarrow) = 2i+1~.
\label{eq:JW_index_mapping}
\end{equation}
Defining the intermediate JW string between sites $i < j$ as $Z^{\mathrm{JW}}_{ij} = \prod_{k=i+1}^{j-1} Z_{\pi(k,\uparrow)} Z_{\pi(k,\downarrow)}$, we apply the transformation to each spin contribution to $\omega^{\braket{i,j}} = \omega^{\braket{i,j}}_\uparrow + \omega^{\braket{i,j}}_\downarrow$, which yields
\begin{widetext}
\begin{align}
\omega^{\braket{i,j}}_{\uparrow}
&=
\frac{1}{8}
\Biggl(
Z_{\pi(i,\downarrow)} - Z_{\pi(j,\downarrow)}
\Biggr)
\Biggl(
X_{\pi(i,\uparrow)} Z_{\pi(i,\downarrow)} Z^{\mathrm{JW}}_{ij} Y_{\pi(j,\uparrow)}
-
Y_{\pi(i,\uparrow)} Z_{\pi(i,\downarrow)} Z^{\mathrm{JW}}_{ij} X_{\pi(j,\uparrow)}
\Biggr)
\label{eq:JW_omega_local_components_up}
\\
\omega^{\braket{i,j}}_{\downarrow}
&=
\frac{1}{8}
\Biggl(
Z_{\pi(i,\uparrow)} - Z_{\pi(j,\uparrow)}
\Biggr)
\Biggl(
X_{\pi(i,\downarrow)} Z^{\mathrm{JW}}_{ij} Z_{\pi(j,\uparrow)} Y_{\pi(j,\downarrow)}
-
Y_{\pi(i,\downarrow)} Z^{\mathrm{JW}}_{ij} Z_{\pi(j,\uparrow)} X_{\pi(j,\downarrow)}
\Biggr)
\label{eq:JW_omega_local_components_down}
~,
\end{align}
\end{widetext}
where $X_{\pi(i,\sigma)}$, $Y_{\pi(i,\sigma)}$, and $Z_{\pi(i,\sigma)}$ act on the qubit assigned to the spin orbital $(i,\sigma)$ according to Eq.~\eqref{eq:JW_index_mapping}.

Since the spin components $\omega^{\braket{i,j}}_{\uparrow}$ and $\omega^{\braket{i,j}}_{\downarrow}$ commute, the unitary $\mathcal{W}^{\braket{i,j}}(\phi)$ factorizes as
\begin{equation}
\mathcal{W}^{\braket{i,j}}(\phi)
=
e^{-i\frac{\phi}{2}\omega^{\braket{i,j}}_{\uparrow}}
e^{-i\frac{\phi}{2}\omega^{\braket{i,j}}_{\downarrow}}~.
\end{equation}
Each component in Eqs.~\eqref{eq:JW_omega_local_components_up} and \eqref{eq:JW_omega_local_components_down} contains four mutually commuting Pauli strings, allowing $\mathcal{W}^{\braket{i,j}}(\phi)$ to be implemented exactly as a product of eight Pauli-string rotation gates.
Each of these eight rotation gates can be implemented in the form
\begin{equation}
e^{-i\frac{\alpha}{2} P} = C_P^{\dagger} R_Z(\alpha) C_P~,
\end{equation}
where $P$ denotes a Pauli string and $C_P$ is a Clifford operator satisfying $C_P^{\dagger} Z C_P = P$.
The $R_Z$ gate can act on any qubit in the support of $P$, with the corresponding choice of $C_P$.
The Pauli strings appearing in Eqs.~\eqref{eq:JW_omega_local_components_up} and~\eqref{eq:JW_omega_local_components_down} act on at most $2(j-i)+2$ qubits for $i<j$.
For nearest-neighbor bonds on the $L_x\times L_y$ square lattice with row-major site ordering ($i = yL_x + x$), the two sites differ by $j-i=1$ along $x$ and $j-i=L_x$ along $y$.
Thus, their supports are bounded by $2L_x+2$ qubits.

\section{Fidelity estimation and ground-state fidelity bounds}\label{sec:direct_fidelity_estimation}

While the circuit construction in Sec.~\ref{sec:variational_quantum_circuits} specifies the architecture, the variational parameters must still be determined.
For small lattices, these parameters can be optimized using full state-vector simulations.
Here, we optimize the parameters on a $4\times4$ lattice and transfer them to larger systems without further optimization.
We then evaluate how well the transferred parameters perform on these targets.

As our primary figure of merit, we consider the ground-state fidelity, which dictates both the success probability
and the cost of subsequent ground-state energy estimation.
In the absence of the exact ground state, we evaluate the fidelity of the circuit state relative to a VMC reference state.
By combining this sampled fidelity and the reference energy with estimates of the ground-state energy and singlet gap, we obtain a lower bound on the circuit's ground-state fidelity.

\subsection{Fidelity to the VMC reference}
\label{sec:vmc_reference_fidelity}

We construct the reference state using the fermion-pair wave function implemented in the Many-Variable Variational Monte Carlo (mVMC) software package~\cite{Misawa2019mVMC}.
This variational ansatz for the reference state is distinct from the circuit ansatz and is optimized directly to approximate the ground state.
For the Heisenberg model, imposing single occupancy restricts the wave function to the spin Hilbert space.
We then apply numerical spin projection to select the $S=0$ subspace while preserving $\SUtwo$ symmetry.
We optimize the variational parameters using stochastic reconfiguration (SR) and solve the resulting linear system with the conjugate gradient (CG) method.
Appendix~\ref{app:vmc_reference} describes the specific form of the VMC ansatz and how its amplitudes are evaluated in the computational basis.

The fidelity between the circuit state $\ket{\Psi(\bm{\theta})}$ and the reference state $\ket{\Psi_{\mathrm{ref}}}$ is defined as
\begin{equation}
F_{\mathrm{ref}}
=
\frac{\left|\braket{\Psi_{\mathrm{ref}}|\Psi(\bm{\theta})}\right|^2}
{\braket{\Psi_{\mathrm{ref}}|\Psi_{\mathrm{ref}}}\braket{\Psi(\bm{\theta})|\Psi(\bm{\theta})}}~.
\label{eq:reference_fidelity}
\end{equation}
We rewrite $F_{\mathrm{ref}}$ in terms of the computational-basis amplitudes $\Psi_{\mathrm{ref}}(\bm{s})=\braket{\bm{s}|\Psi_{\mathrm{ref}}}$ and $\Psi(\bm{s};\bm{\theta})=\braket{\bm{s}|\Psi(\bm{\theta})}$, where $\bm{s}$ runs over all spin configurations.
Assuming that $\Psi(\bm{s};\bm{\theta})$ can be evaluated for any $\bm{s}$ by classically simulating the circuit, we define the amplitude ratio
\begin{equation}
R(\bm{s})
=
\frac{\Psi(\bm{s};\bm{\theta})}
{\Psi_{\mathrm{ref}}(\bm{s})}~,
\label{eq:amplitude_ratio}
\end{equation}
and assume that the circuit state has no support on configurations for which the reference amplitude vanishes, so that $\Psi(\bm{s};\bm{\theta})=\Psi_{\mathrm{ref}}(\bm{s})R(\bm{s})$ holds for every $\bm{s}$.
Substituting $\Psi(\bm{s};\bm{\theta})=\Psi_{\mathrm{ref}}(\bm{s})R(\bm{s})$ into Eq.~\eqref{eq:reference_fidelity} gives
\begin{equation}
F_{\mathrm{ref}}
=
\frac{\left|\sum_{\bm{s}}|\Psi_{\mathrm{ref}}(\bm{s})|^2R(\bm{s})\right|^2}
{\left(\sum_{\bm{s}}|\Psi_{\mathrm{ref}}(\bm{s})|^2\right)
\left(\sum_{\bm{s}}|\Psi_{\mathrm{ref}}(\bm{s})|^2|R(\bm{s})|^2\right)}~.
\label{eq:reference_fidelity_exact}
\end{equation}

Dividing each sum by $\braket{\Psi_{\mathrm{ref}}|\Psi_{\mathrm{ref}}}=\sum_{\bm{s}}|\Psi_{\mathrm{ref}}(\bm{s})|^2$ turns it into an expectation value under the reference distribution
\begin{equation}
p(\bm{s})
=
\frac{|\Psi_{\mathrm{ref}}(\bm{s})|^2}
{\braket{\Psi_{\mathrm{ref}}|\Psi_{\mathrm{ref}}}}~,
\label{eq:reference_distribution}
\end{equation}
so that
\begin{equation}
F_{\mathrm{ref}}
=
\frac{|\braket{R}_{p}|^2}
{\braket{\lvert R\rvert^2}_{p}}~,
\label{eq:reference_fidelity_estimator}
\end{equation}
where $\braket{h}_p=\sum_{\bm{s}}p(\bm{s})h(\bm{s})$.
The expectation values in Eq.~\eqref{eq:reference_fidelity_estimator} run over exponentially many configurations.
We therefore estimate them by sample averages over configurations drawn from $p(\bm{s})$ with the Metropolis--Hastings algorithm~\cite{Metropolis1953,Hastings1970}.
Since both the circuit and the reference state conserve $\Stot^z$, the sampling is restricted to the zero-magnetization sector $N_\uparrow=N_\downarrow=N/2$, outside of which both amplitudes vanish.

\subsection{Inferring fidelity to the exact ground state}
\label{sec:ground_fidelity_bound}

The fidelity $F_{\mathrm{ref}}$ in Eq.~\eqref{eq:reference_fidelity} is defined with respect to the VMC reference state, rather than the exact ground state.
To infer the circuit's ground-state fidelity from this quantity, we first bound the fidelity of the reference state with respect to the exact ground state.
For the Heisenberg model, we use the reference energy, a ground-state energy estimate from stochastic series expansion (SSE) quantum Monte Carlo~\cite{Sandvik1997}, and an estimate of the excitation gap in the corresponding spin subspace.

As discussed in Sec.~\ref{sec:variational_ansatz}, the ground state of the open square lattice studied here, with an even number of sites, is a unique singlet.
The eSWAP circuit preserves the $S=0$ subspace, while the VMC reference is numerically projected onto this subspace.
Assuming that the numerical projection sufficiently removes contributions from $S>0$ components, only singlet excitations contribute to the bound below.
The relevant energy gap is therefore
\begin{equation}
\Delta_0=E_1^{(0)}-E_0^{(0)},\qquad E_0\equiv E_0^{(0)},
\label{eq:singlet_gap_definition}
\end{equation}
where $E_0^{(0)}$ and $E_1^{(0)}$ are the two lowest eigenenergies in the $S=0$ subspace.

This gap is distinct from the singlet--triplet gap.
For $L_x=L_y=L$, the low-lying $S>0$ states in the Anderson tower of states follow the energy scaling
\begin{equation}
E_0^{(S)}-E_0^{(0)}\propto\frac{S(S+1)}{L^2},
\label{eq:tower_of_states_gap}
\end{equation}
where $E_0^{(S)}$ denotes the ground-state energy in the subspace with total spin $S$~\cite{Anderson1952,HasenfratzNiedermayer1993}.
As these states have no overlap with the singlet reference, they do not enter the fidelity bound.

\begin{figure}[tbp]
\centering
\includegraphics[clip,width=1.0\linewidth]{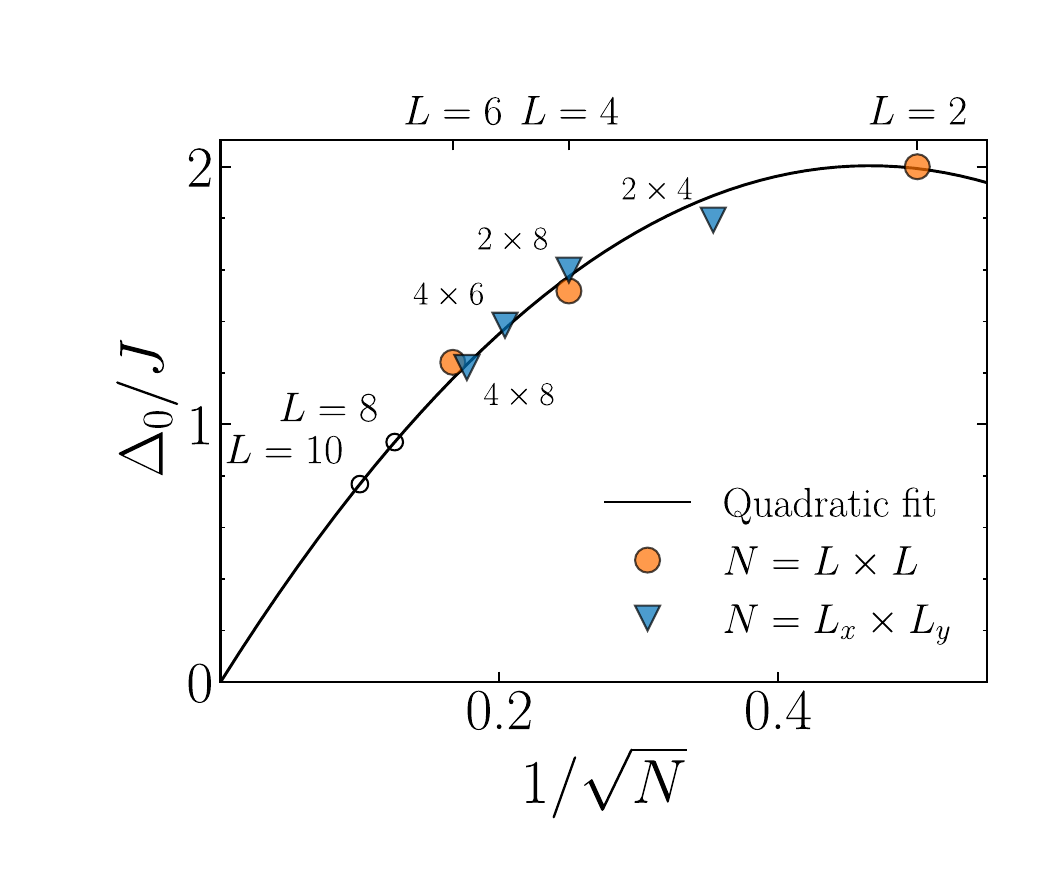}
\caption{Singlet gap $\Delta_0/J$ of the open-boundary Heisenberg model.
Filled circles are exact-diagonalization data fitted to a quadratic in $1/L$ with zero intercept; open circles are extrapolations.
Triangles show rectangular-cluster data excluded from the fit.}
\label{fig:gap_scaling_heis}
\end{figure}

In addition to the tower of states, the ordered antiferromagnetic phase hosts long-wavelength spin-wave excitations whose finite-size energies scale as $1/L$~\cite{NeubergerZiman1989,WeihongHamer1993}.
Since two spin-1 magnons can combine into a singlet, the lowest singlet excitation should follow the same leading finite-size scaling.
We therefore perform exact diagonalization on $L\times L$ open square clusters with $L=2,4,6$, and fit the resulting singlet gaps to
\begin{equation}
\frac{\Delta_0(L)}{J}=\frac{a_1}{L}+\frac{a_2}{L^2}~,
\label{eq:singlet_gap_extrapolation}
\end{equation}
imposing a vanishing gap in the thermodynamic limit.
Here, $a_1$ and $a_2$ are fitting parameters that include boundary corrections.
The resulting fit yields $a_1=8.6074$ and $a_2=-9.2429$, as shown in Fig.~\ref{fig:gap_scaling_heis}.
While the rectangular-cluster data are not included in the fit, they also agree well with the fitted curve when plotted against $1/\sqrt{N}$.
For our subsequent analysis, we use the exact gaps for $L=4$ and $6$, and extrapolate to larger systems using Eq.~\eqref{eq:singlet_gap_extrapolation}.
Note that all exact diagonalizations in this work are performed using the XDiag package~\cite{Wietek_2026}.

We next relate the reference energy to the ground-state fidelity via the energy gap.
For normalized states $\ket{\Psi(\bm{\theta})}$ and $\ket{\Psi_{\mathrm{ref}}}$, their fidelities with respect to the exact ground state $\ket{\Psi_0}$ are defined as
\begin{align}
F_0&=\lvert\braket{\Psi_0|\Psi(\bm{\theta})}\rvert^2,\\
F_{\mathrm{ref},0}&=\lvert\braket{\Psi_0|\Psi_{\mathrm{ref}}}\rvert^2.
\end{align}
Letting $E_{\mathrm{ref}}$ denote the energy of the reference state, we expand $\ket{\Psi_{\mathrm{ref}}}$ in the eigenbasis of the $S=0$ subspace.
Combining this expansion with $E_0$ and $\Delta_0$ from Eq.~\eqref{eq:singlet_gap_definition} yields the lower bound
\begin{equation}
F_{\mathrm{ref},0}\geq F_{\mathrm{ref},0}^{\min}
\equiv\left[1-\frac{E_{\mathrm{ref}}-E_0}{\Delta_0}\right]_+~,
\label{eq:vmc_energy_fidelity_bound}
\end{equation}
where $[x]_+=\max\{x,0\}$.
This lower bound is rigorous when evaluated with the true ground-state energy and gap.
Using the SSE ground-state energy and the singlet gap obtained above, we evaluate Eq.~\eqref{eq:vmc_energy_fidelity_bound} for the VMC states listed in Table~\ref{tab:vmc_ground_bound}.
Defining the energy error as $|\Delta E|=|E_{\mathrm{ref}}-E_0|$, the resulting lower bounds exceed $0.995$ for $L=4, 6$, and $8$, and remain above $0.9564$ even for $L=10$.

\begin{table}[t]
\centering
\caption{Heisenberg VMC energy errors relative to SSE, singlet gaps, and fidelity lower-bound estimates.
Gaps for $L=8,10$ are extrapolated.
}
\label{tab:vmc_ground_bound}
\begin{tabular}{c|ccc}
\hline
$L$ & $|\Delta E|/J$ & $\Delta_0/J$ & $F_{\mathrm{ref},0}^{\min}$ \\
\hline
4 & $6.221\times10^{-4}$ & $1.518$ & $0.9996$ \\
6 & $4.823\times10^{-3}$ & $1.241$ & $0.9961$ \\
8 & $4.038\times10^{-3}$ & $0.9315$ & $0.9957$ \\
10 & $3.346\times10^{-2}$ & $0.7683$ & $0.9564$ \\
\hline
\end{tabular}
\end{table}

We use the reference bound $F_{\mathrm{ref},0}^{\min}$ and the reference fidelity $F_{\mathrm{ref}}$ to obtain a lower bound on $F_0$.
For two normalized pure states, the Fubini--Study angle is defined as $\vartheta(\psi,\phi)=\arccos\lvert\braket{\psi|\phi}\rvert$, which satisfies the triangle inequality.
Applying this inequality to the circuit state, the reference state, and the exact ground state, and using $F_{\mathrm{ref},0}\geq F_{\mathrm{ref},0}^{\min}$, we obtain
\begin{equation}
\begin{split}
F_0\geq F_0^{\min}\equiv
\Biggl[&\sqrt{F_{\mathrm{ref}}F_{\mathrm{ref},0}^{\min}}\\
&-\sqrt{(1-F_{\mathrm{ref}})(1-F_{\mathrm{ref},0}^{\min})}\Biggr]_{+}^{2}.
\end{split}
\label{eq:circuit_ground_fidelity_bound}
\end{equation}
This lower bound is positive when $F_{\mathrm{ref}}+F_{\mathrm{ref},0}^{\min}>1$ and is zero otherwise.
To rigorously apply this bound to the circuit state, one must also take into account the simulation and statistical errors in $F_{\mathrm{ref}}$.

\section{Circuit benchmarks and preparation costs}
\label{sec:results}

We first describe the tensor-network method used to simulate the circuit states.
Next, we validate the fidelity estimates on a $4\times4$ lattice and evaluate parameter transferability to larger Heisenberg lattices.
We then optimize the SW charge-fluctuation gates and evaluate the Hubbard ground-state fidelity on the $4\times4$ lattice.
Finally, we assess the state-preparation overhead for both models.

\subsection{Tensor-network simulation}
\label{sec:Classical_optimization_of_VQC}

\begin{figure}[t]
\centering
\includegraphics[clip,width=0.9\linewidth]{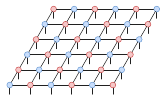}
\caption{PEPS on a square lattice.
Circles denote site tensors, connecting lines denote virtual indices of bond dimension $D$, and open legs denote physical spin indices of dimension $d=2$.}
\label{fig:PEPS}
\end{figure}

To evaluate the Heisenberg circuit amplitudes in Sec.~\ref{sec:direct_fidelity_estimation}, we represent the circuit state as a projected entangled-pair state (PEPS), as illustrated in Fig.~\ref{fig:PEPS}.
On an $L_x\times L_y$ square lattice, each site tensor $T_i$ has a physical index of dimension $d=2$ and up to four virtual indices of bond dimension $D$ connecting neighboring sites.
The bond dimension $D$ controls the entanglement retained in the tensor-network representation.
The singlet-dimer state in Eq.~\eqref{eq:dimer_products} has a PEPS bond dimension of $D=2$ on $\mathcal B_{\mathrm x_0}$ and $D=1$ elsewhere.
Each eSWAP gate increases the corresponding bond dimension by at most a factor of $4$, giving $D_{\max}=2\times4^{N_{\mathrm b}}$ without truncation.

\begin{figure}[t]
\centering
\includegraphics[clip,width=0.9\linewidth]{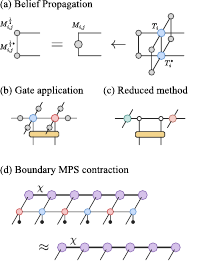}
\caption{PEPS simulation and amplitude evaluation.
(a)~The message $M_{i,j}$ is obtained by contracting the site tensor and its conjugate with the incoming messages.
(b)~A two-site gate is applied and the updated bond is truncated using the converged BP messages as an approximate environment.
(c)~The square roots of the BP messages are first absorbed into the site tensors, which are then decomposed by QR.
The gate application and SVD are performed on the resulting smaller tensors.
(d)~With the physical indices fixed to a sampled configuration, each row is absorbed into a boundary MPS and compressed to bond dimension $\chi$.}
\label{fig:TNSimulations}
\end{figure}

We implement gate operations via the belief propagation (BP)-guided simple update~\cite{Alkabetz2021BP,Tindall2023Gauging}, following recent approaches in large-scale circuit simulation~\cite{Tindall2024Eagle,Patra2024IBM,Begusic2024Fast,Tindall2026BP}.
As sketched in Fig.~\ref{fig:TNSimulations}(a), an approximate environment is constructed before each gate layer by iterating the BP messages to convergence.
Applying a gate expands the bond dimension, which we then truncate back to $D$ through a singular value decomposition (SVD), as shown in Fig.~\ref{fig:TNSimulations}(b).
To keep the algorithm computationally efficient, the square roots of the converged messages are absorbed into the site tensors, followed by a QR decomposition.
As depicted in Fig.~\ref{fig:TNSimulations}(c), this procedure allows both the gate application and the SVD truncation to proceed on reduced tensors~\cite{Phien2015Reduced}, restricting the leading computational cost to $\mathcal O(D^5)$ per local update.

Given a sampled configuration $\bm s$, we fix the physical indices to obtain a single-layer network that contracts to $\Psi(\bm s;\bm\theta)$.
We contract this network row by row using a boundary MPS, compressing the boundary MPS to bond dimension $\chi$ after each row via variational fitting~\cite{Vieijra2021DirectSampling}, as illustrated in Fig.~\ref{fig:TNSimulations}(d).
The contraction cost scales as $\mathcal O(\chi^2D^4+\chi^3D^2)$.
In practice, $\chi=\mathcal O(D)$ is sufficient~\cite{PhysRevB.103.235155,r4q9-4yvj}, giving an effective scaling of $\mathcal O(D^6)$.
Amplitudes for different configurations are evaluated in parallel batches on multiple GPUs~\cite{RudolphTindall2025}.

\subsection{Heisenberg parameter transfer}
\label{sec:heisenberg_transfer}

We evaluate the Heisenberg circuits defined in Eq.~\eqref{eq:variational_state} with $N_{\mathrm{b}} \in \{2, 3, 4\}$ blocks on $L \times L$ square lattices.
For the parameter-transfer benchmarks, the variational parameters are shared across all gates within each layer, and the resulting angles are applied directly to larger lattices without further optimization.

\begin{figure}[t]
\centering
\includegraphics[clip,width=1.0\linewidth]{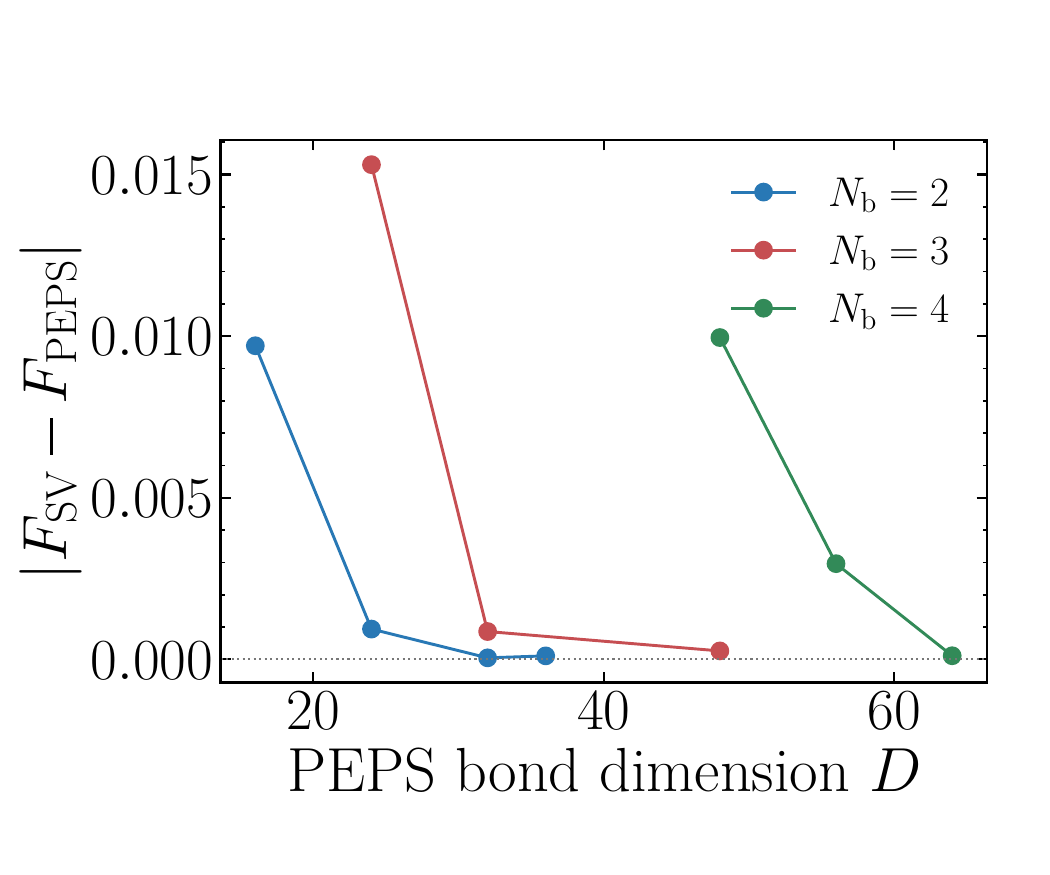}
\caption{Bond-dimension dependence of the ground-state fidelity error $\lvert F_{\mathrm{SV}}-F_{\mathrm{PEPS}}\rvert$ for the $4\times4$ lattice, with $\chi=3D$.
Blue, red, and green circles represent $N_{\mathrm b}=2$, $3$, and $4$, respectively.
}
\label{fig:fidelity_l4_vs_D}
\end{figure}

To benchmark the accuracy of our PEPS-based fidelity evaluations, we first compare them with exact state-vector calculations on a $4\times4$ lattice.
Here, $F_{\mathrm{PEPS}}$ denotes the fidelity evaluated via PEPS, while the exact fidelity $F_{\mathrm{SV}}$ between the circuit state and the ground state is $0.8253$ for $N_{\mathrm b}=2$, $0.9539$ for $N_{\mathrm b}=3$, and $0.9774$ for $N_{\mathrm b}=4$.
We vary the PEPS bond dimension $D$ while keeping $\chi=3D$ when contracting the amplitudes.
To compute the necessary inner products, we sum over the $\binom{16}{8}=12{,}870$ configurations in the $S_{\mathrm{tot}}^z=0$ subspace and evaluate the log-amplitude contractions in single precision.
Figure~\ref{fig:fidelity_l4_vs_D} shows that $F_{\mathrm{PEPS}}$ converges to $F_{\mathrm{SV}}$ as $D$ increases for all three values of $N_{\mathrm b}$.
In the subsequent parameter-transfer calculations, we use $D=36$ for $N_{\mathrm b}=2$, $D=48$ for $N_{\mathrm b}=3$, and $D=64$ for $N_{\mathrm b}=4$, with $\chi=3D$.

Next, we transfer the parameters optimized on the $4\times4$ lattice to larger systems without further optimization and estimate the reference fidelity $F_{\mathrm{ref}}$ in Eq.~\eqref{eq:reference_fidelity_estimator} using 20,000 samples from each VMC reference state.
Table~\ref{tab:transferred_fidelity} lists the reference fidelities and the ground-state fidelity lower-bound estimates $F_0^{\min}$ obtained using Eq.~\eqref{eq:circuit_ground_fidelity_bound}.
Increasing $N_{\mathrm b}$ improves the reference fidelity over the range studied.
Notably, for $N_{\mathrm b}=3$, both the reference fidelity and the estimated lower bound on the ground-state fidelity remain nonzero for lattices up to $10\times10$.
We again emphasize that $F_0^{\min}$ is a lower-bound estimate, so the actual ground-state fidelity can be higher.

Table~\ref{tab:transferred_fidelity} also reports the estimated lower bound on the overlap amplitude, $\sqrt{F_0^{\min}}$.
Ground-state preparation algorithms based on amplitude amplification scale with the initial amplitude as $1/\sqrt{F_0}$, offering a quadratic speedup over the $1/F_0$ scaling required by QPE~\cite{ge2018fastergroundstatepreparation,Lin2020nearoptimalground}.
Because this lower bound remains non-negligible even for the $10\times10$ lattice with $N_{\mathrm b}=3$, the associated overhead is expected to remain modest.

\begin{table}[tbp]
\centering
\caption{Reference fidelities and ground-state fidelity lower-bound estimates for Heisenberg circuits transferred from the $4\times4$ lattice.}
\label{tab:transferred_fidelity}

\begin{tabular}{c|c|c|c|c}
\hline
$L$ & $N_{\mathrm b}$ & $F_{\mathrm{ref}}$ & $F_0^{\min}$ & $\sqrt{F_0^{\min}}$ \\
\hline
\multirow{3}{*}{6} & $2$ & $0.4532 \pm 0.0030$ & $0.3915$ & $0.6257$ \\
& $3$ & $0.7055 \pm 0.0026$ & $0.6471$ & $0.8044$ \\
& $4$ & $0.7903 \pm 0.0022$ & $0.7373$ & $0.8587$ \\
\hline
\multirow{3}{*}{8} & $2$ & $0.1594 \pm 0.0035$ & $0.1144$ & $0.3382$ \\
& $3$ & $0.3921 \pm 0.0033$ & $0.3291$ & $0.5737$ \\
& $4$ & $0.4975 \pm 0.0032$ & $0.4321$ & $0.6573$ \\
\hline
\multirow{3}{*}{10} & $2$ & $0.0337 \pm 0.0023$ & $0$ & $0$ \\
& $3$ & $0.1368 \pm 0.0028$ & $0.0281$ & $0.1676$ \\
& $4$ & $0.2065 \pm 0.0033$ & $0.0668$ & $0.2584$ \\
\hline
\end{tabular}
\end{table}

\subsection{Hubbard circuit parameters and exact fidelity}
\label{sec:hubbard_preparation}

Having established parameter transferability for the Heisenberg circuit, we now incorporate the charge-fluctuation block defined in Eq.~\eqref{eq:charge_block} to construct the Hubbard circuit introduced in Eq.~\eqref{eq:hubbard_variational_state}.
On a $4\times4$ lattice, we keep the layer-shared parameters of the Heisenberg circuit in Eq.~\eqref{eq:variational_state} fixed at the values used in Sec.~\ref{sec:heisenberg_transfer} for $N_{\mathrm b}=3,4,5$ and optimize only the four layer-shared charge-fluctuation angles $\bm{\phi}$ against the exact Hubbard ground state.
These angles are initialized to $\phi=4t/U$ across all four layers, consistent with the first-order SW transformation at $U/t=8$ (Sec.~\ref{sec:variational_ansatz}).
We evaluate the resulting fidelities via state-vector simulations.

Tables~\ref{tab:hubbard_exact} and~\ref{tab:hubbard_angles} show the fidelities and the optimized angles, respectively.
In Table~\ref{tab:hubbard_exact}, ``No SW block'' refers to the Heisenberg circuit state embedded in the singly occupied fermionic subspace using Eq.~\eqref{eq:V_mapping}, without any charge-fluctuation gates, i.e., $\bm \phi = \bm 0$.
Adding the charge block with the SW angles improves the fidelity relative to the exact Hubbard ground state for all studied values of $N_{\mathrm b}$.
Optimizing the charge angles leads to a further improvement.
The fidelity increases with $N_{\mathrm b}$, while the optimized charge angles remain nearly unchanged as $N_{\mathrm b}$ varies (see Table~\ref{tab:hubbard_angles}).
This suggests that the improvement mainly comes from the quality of the underlying Heisenberg state.

\begin{table}[tbp]
\centering
\caption{Hubbard ground-state fidelities on the $4\times4$ lattice at $U/t=8$ for the three parameter settings, with fixed Heisenberg parameters.}
\label{tab:hubbard_exact}
\begin{tabular}{c|ccc}
\hline
$N_{\mathrm b}$ & No SW block & $\phi = 4t/U$ & Optimized $\phi$ \\
\hline
3 & 0.4212 & 0.8343 & 0.8515 \\
4 & 0.4419 & 0.8763 & 0.8941 \\
5 & 0.4522 & 0.8971 & 0.9152 \\
\hline
\end{tabular}
\end{table}

\begin{table}[tbp]
\centering
\caption{Optimized charge-gate angles (radians) for the $4\times4$ Hubbard circuit at $U/t=8$, listed in application order.}
\label{tab:hubbard_angles}
\begin{tabular}{c|ccc}
\hline
Matching & $N_{\mathrm b}=3$ & $N_{\mathrm b}=4$ & $N_{\mathrm b}=5$ \\
\hline
$\mathcal B_{\mathrm y_1}$ & 0.4235 & 0.4235 & 0.4233 \\
$\mathcal B_{\mathrm y_0}$ & 0.4249 & 0.4255 & 0.4259 \\
$\mathcal B_{\mathrm x_1}$ & 0.4278 & 0.4280 & 0.4277 \\
$\mathcal B_{\mathrm x_0}$ & 0.4276 & 0.4284 & 0.4289 \\
\hline
\end{tabular}
\end{table}

These parameters, which are shared across layers for the Heisenberg and charge-fluctuation terms, can be reused to construct Hubbard circuits for larger systems without further optimization.
The successful transfer of parameters in the Heisenberg model suggests that this strategy can be readily extended to prepare Hubbard ground states on larger lattices.

As $U/t$ increases, the Hubbard ground state approaches the Heisenberg limit, so the circuit fidelities are expected to approach those obtained for the Heisenberg model in Sec.~\ref{sec:heisenberg_transfer}. 
A natural extension would be to use a Hamiltonian variational ansatz~\cite{PhysRevA.92.042303,Park2024hamiltonian} with deeper circuits. Optimizing the parameters of these circuits could further improve their fidelities for larger lattices, as indicated by our Heisenberg results.

\subsection{Circuit resources}

We evaluate the state-preparation cost by counting the number of rotations in the circuit detailed in Sec.~\ref{sec:quantum_circuit_construction}.
The decomposition in Eq.~\eqref{eq:eswap_decomposition} uses three Pauli rotations per eSWAP gate, while Eqs.~\eqref{eq:JW_omega_local_components_up} and~\eqref{eq:JW_omega_local_components_down} give eight rotations per charge-fluctuation gate.
Each rotation is equivalent to an $R_Z$ rotation up to Clifford gates.
The full Hubbard circuit in Eq.~\eqref{eq:hubbard_variational_state} therefore requires
\begin{equation}
\begin{split}
n_{R_Z}
&=3N_{\mathrm b}|\mathcal B|+8|\mathcal B|\\
&=(6N_{\mathrm b}+16)L^2-(6N_{\mathrm b}+16)L~,
\end{split}
\label{eq:rotation_counts}
\end{equation}
where $|\mathcal B|=2L(L-1)$ is the number of nearest-neighbor bonds on the open $L\times L$ square lattice.
This count reflects the circuit prior to simplification and is unaffected by parameter sharing.
Note that singlet preparation, embedding, and JW strings introduce only Clifford gates.

For Clifford+$T$ synthesis with an operator-norm error of $\epsilon$ per rotation, the leading-order $T$-count estimate follows Ref.~\cite{RossSelinger2016}:
\begin{equation}
n_T\simeq 3n_{R_Z}\log_2(1/\epsilon).
\label{eq:preparation_t_count}
\end{equation}
For a fixed number of blocks $N_{\mathrm b}$ and target accuracy per rotation, this cost scales linearly with the number of lattice bonds.

Table~\ref{tab:t_resources} compares the estimated $T$ counts for lattice sizes from $4\times4$ to $10\times10$ at $\epsilon=10^{-5}$, showing a $7.5$-fold increase across this range.
Each Heisenberg block requires approximately $9|\mathcal B|\log_2(1/\epsilon)$ $T$ gates.
Incorporating the charge-fluctuation block into the $N_{\mathrm b}$-block Heisenberg circuit introduces an additional $24|\mathcal B|\log_2(1/\epsilon)$ $T$ gates, as derived from Eqs.~\eqref{eq:rotation_counts} and~\eqref{eq:preparation_t_count}.
Combined with the fidelity results in Table~\ref{tab:transferred_fidelity}, these figures quantify the cost of enhancing Heisenberg ground-state preparation via larger $N_{\mathrm b}$.
Across both models, the preparation overhead remains on the order of $10^5$ $T$ gates for the reported sizes and depths, demonstrating the feasibility of using these states as inputs for subsequent ground-state energy estimation.

\begin{table}[tbp]
\centering
\caption{Estimated $T$ counts for the Heisenberg and Hubbard circuits, using 50 $T$ gates per rotation at $\epsilon=10^{-5}$.}
\label{tab:t_resources}
{
\begin{tabular}{cc|rr}
\hline
$L$ & $N_{\mathrm b}$ & Heisenberg & Hubbard \\
\hline
4 & 3 & $1.08\times10^4$ & $2.04\times10^4$ \\
4 & 4 & $1.44\times10^4$ & $2.40\times10^4$ \\
4 & 5 & $1.80\times10^4$ & $2.76\times10^4$ \\
\hline
6 & 3 & $2.70\times10^4$ & $5.10\times10^4$ \\
6 & 4 & $3.60\times10^4$ & $6.00\times10^4$ \\
6 & 5 & $4.50\times10^4$ & $6.90\times10^4$ \\
\hline
8 & 3 & $5.04\times10^4$ & $9.52\times10^4$ \\
8 & 4 & $6.72\times10^4$ & $1.12\times10^5$ \\
8 & 5 & $8.40\times10^4$ & $1.29\times10^5$ \\
\hline
10 & 3 & $8.10\times10^4$ & $1.53\times10^5$ \\
10 & 4 & $1.08\times10^5$ & $1.80\times10^5$ \\
10 & 5 & $1.35\times10^5$ & $2.07\times10^5$ \\
\hline
\end{tabular}}
\end{table}

\section{Conclusion}
\label{sec:conclusion}

We construct ground-state preparation circuits for the Hubbard model by combining a Heisenberg circuit with an SW charge-fluctuation block.
The Heisenberg benchmarks indicate that parameters optimized on a small lattice remain effective as the lattice size increases.
For the $N_{\mathrm b}=3$ Heisenberg circuit, the ground-state fidelity lower bounds remain positive across all studied system sizes.
Increasing the circuit depth also improves the fidelity to the target state over the range studied.
When classical simulation is feasible, directly optimizing the rotation angles can further improve the fidelity without increasing the gate count (Appendix~\ref{sec:direct_fidelity_optimization}).

Adding the charge-fluctuation block improves the circuit fidelity to the Hubbard ground state in exact calculations for small lattices.
The resulting parameters define a concrete ground-state preparation circuit for a $10\times10$ lattice, with an estimated preparation cost of approximately $10^5$ $T$ gates.
This cost scales linearly with the number of lattice bonds when the number of blocks and the rotation precision are held fixed.
Taken together, our fidelity benchmarks and circuit-resource estimates indicate that useful initial states can be prepared with resource costs orders of magnitude below those reported for subsequent QPE-based ground-state energy estimation.
The layer-shared circuits in OpenQASM format, their parameter values, and the export code are provided in our GitHub repository~\cite{GroundStatePreparationCircuits}.

Estimating fidelity lower bounds for larger Hubbard systems will require not only reliable estimates of the relevant energy scales and spectral gaps, but also, crucially, accurate circuit simulations.
Currently, BP truncation limits the reach of our approach by demanding large bond dimensions~\cite{Midha2026BP,Bermejo2026BPLimits}, despite the modest circuit entanglement observed in the regime accessible via exact simulation.
More accurate, albeit computationally demanding, alternatives include full-update schemes that incorporate the norm environment of the entire system~\cite{PhysRevLett.101.250602,Lubasch_2014}, as well as variational optimization via VMC~\cite{WuNys2025TimeDependentVMC,DuChan2025Neuralized}.
For more challenging models, expressive ansatzes such as neural quantum states and VMC-PEPS can still provide valuable fidelity benchmarks, even in the absence of rigorous bounds on the ground-state fidelity~\cite{CarleoTroyer2017,Nomura2017RBM}.

\section*{Acknowledgments}
This work is supported by MEXT Quantum Leap Flagship Program (MEXT Q-LEAP) Grant No.~JPMXS0120319794, JST COI-NEXT Grant No.~JPMJPF2014, and JST CREST Grant No.~JPMJCR24I3.
R.W. also acknowledges support from JSPS KAKENHI Grant No.~25KJ1773 and JST CREST Grant No.~JPMJCR24I1.

\bibliography{main}

\appendix

\section{Definition of the charge-fluctuation gate}\label{sec:SW_transformation}

Let $\Omega$ be an anti-Hermitian operator and consider the unitary transformation
\begin{equation}
e^{\Omega} H e^{-\Omega}
=
H + [\Omega,H] + \frac{1}{2}[\Omega,[\Omega,H]] + \cdots~,
\label{eq:SW_expansion}
\end{equation}
where the right-hand side is expanded using the Baker--Campbell--Hausdorff formula.
The hopping terms that connect states with different doublon numbers are
\begin{equation}
H_{\mathrm{off}}
=
-t \sum_{\braket{i,j}\in\mathcal{B},\sigma}
\bigl(n_{i,\bar\sigma}-n_{j,\bar\sigma}\bigr)^2
\bigl(
c^\dagger_{i,\sigma} c_{j,\sigma}
+
c^\dagger_{j,\sigma} c_{i,\sigma}
\bigr)~,
\label{eq:SW_off_diagonal_terms}
\end{equation}
where $\bar\sigma$ denotes the spin opposite to $\sigma$.
To eliminate these off-diagonal terms from the transformed Hamiltonian up to first order in $t/U$, we choose the leading-order SW generator $\Omega$ such that
\begin{equation}
H_{\mathrm{off}} + \biggl[\Omega, U\sum_{i\in\Lambda} n_{i,\uparrow} n_{i,\downarrow}\biggr]=0~,
\label{eq:SW_condition}
\end{equation}
which is satisfied by
\begin{equation}
\Omega
=
-\frac{t}{U}
\sum_{\braket{i,j}\in\mathcal{B},\sigma}
\bigl(n_{i,\bar\sigma}-n_{j,\bar\sigma}\bigr)
\bigl(
c^\dagger_{i,\sigma} c_{j,\sigma}
-
c^\dagger_{j,\sigma} c_{i,\sigma}
\bigr)~.
\label{eq:SW_generator_derived}
\end{equation}
Let $\mathcal P$ project onto the singly occupied subspace.
At half filling, the second-order effective Hamiltonian is
\begin{equation}
\begin{split}
\mathcal P e^{\Omega}H e^{-\Omega}\mathcal P
={}&\frac{4t^2}{U}\mathcal P\sum_{\braket{i,j}\in\mathcal B}
\left(\mathbf S_i\cdot\mathbf S_j-\frac14\right)\mathcal P\\
&+\mathcal O\!\left(\frac{t^3}{U^2}\right).
\end{split}
\label{eq:SW_second_order}
\end{equation}
Virtual hopping into and out of doubly occupied configurations generates the antiferromagnetic exchange $J=4t^2/U$.
The projection is essential: hopping that preserves doublon number is present outside this subspace.

As the generator $\Omega$ is anti-Hermitian, we factor out the local Hermitian operator $\omega^{\braket{i,j}}$ defined in Eq.~\eqref{eq:omega_local}, so that
\begin{equation}
\Omega = +2i\frac{t}{U}\sum_{\braket{i,j}\in\mathcal{B}} \omega^{\braket{i,j}}~.
\label{eq:SW_generator_omega}
\end{equation}
Since terms in Eq.~\eqref{eq:SW_generator_omega} on overlapping bonds do not generally commute, the global unitary $e^{-\Omega}$ is split by the Suzuki--Trotter decomposition into the four matchings, which yields the two-site gate $\mathcal{W}^{\braket{i,j}}(\phi)$ defined in Eq.~\eqref{eq:charge_gate} with the SW angle $\phi = 4t/U$ on every layer, as described in Sec.~\ref{sec:variational_ansatz}.

To illustrate the physical action of this local gate, we consider the singlet $\ket{\mathrm{s}_{ij}}$ in Eq.~\eqref{eq:dimer_products} in the fermionic occupation-number basis, together with the doublon--holon state on the same bond:
\begin{equation}
\ket{\mathrm{dh}_{ij}}
=
\frac{1}{\sqrt2}
\bigl(
\ket{\uparrow_i \downarrow_i,0}
+
\ket{0, \uparrow_j \downarrow_j}
\bigr)~.
\end{equation}
Within the subspace spanned by \(\{\ket{\mathrm{s}_{ij}},\ket{\mathrm{dh}_{ij}}\}\), the charge-fluctuation gate acts as
\begin{equation}
\begin{aligned}
\mathcal{W}^{\braket{i,j}}(\phi)\ket{\mathrm{s}_{ij}}
&=
\cos\frac{\phi}{2}\,\ket{\mathrm{s}_{ij}}
+
\sin\frac{\phi}{2}\,\ket{\mathrm{dh}_{ij}}~,
\\
\mathcal{W}^{\braket{i,j}}(\phi)\ket{\mathrm{dh}_{ij}}
&=
\cos\frac{\phi}{2}\,\ket{\mathrm{dh}_{ij}}
-
\sin\frac{\phi}{2}\,\ket{\mathrm{s}_{ij}}~.
\end{aligned}
\end{equation}

\section{VMC reference wave functions and amplitude conventions}
\label{app:vmc_reference}

To obtain the VMC reference state, we employ the mVMC software package~\cite{Misawa2019mVMC}.
The spins are represented in terms of auxiliary fermions, from which the reference state is constructed as the following pair-product state of $N/2$ up-down fermion pairs:
\begin{equation}
\ket{\phi_{\mathrm{pair}}}
=\left(\sum_{i,j} f_{ij}
c_{i,\uparrow}^{\dagger}c_{j,\downarrow}^{\dagger}
\right)^{N/2}\ket{0}.
\label{eq:mvmc_pair_product}
\end{equation}
Here, $f_{ij}$ are variational pair amplitudes, and $\ket{0}$ is the auxiliary-fermion vacuum.
To enforce $S_{\mathrm{tot}}^z=0$, we fix $N_\uparrow=N_\downarrow=N/2$ and impose the single-occupancy condition described below.
The wave function is left unnormalized.

To project onto the $S=0$ subspace discussed in Sec.~\ref{sec:variational_ansatz}, we perform numerical spin projection via an average over global spin rotations, as implemented in mVMC.
Denoting this approximate projector by $\widetilde{\mathcal P}_0$, we apply it consistently to both reference configuration sampling and amplitude evaluation.
Imposing the single-occupancy constraint yields the reference state
\begin{equation}
\ket{\psi_{\mathrm{ref}}}
=\mathcal P_{\mathrm{single}}\widetilde{\mathcal P}_0
\ket{\phi_{\mathrm{pair}}},
\label{eq:mvmc_heis_ansatz}
\end{equation}
where $\mathcal P_{\mathrm{single}}$ restricts the state to the spin Hilbert space by enforcing one auxiliary fermion per site.

For each sampled configuration, the amplitude of the spin-rotated pair-product state is evaluated as the Pfaffian of the pairing matrix restricted to the occupied orbitals.
Summing these amplitudes over the spin-projection quadrature yields $\psi_{\mathrm{ref}}(\bm{s})$, up to a configuration-independent normalization factor.
In the mVMC convention, occupied spin-up orbitals precede occupied spin-down orbitals, with sites ordered separately within each spin sector.
This convention differs from the site-by-site spin-basis ordering defined in Sec.~\ref{sec:variational_ansatz} and used for the circuit amplitude in Eq.~\eqref{eq:amplitude_ratio}.
If we denote by $\tau_{\bm{s}}$ the permutation that maps the occupied orbitals from the mVMC ordering to the site-by-site ordering, the fermionic anticommutation relations introduce a sign factor $\operatorname{sgn}(\tau_{\bm{s}})$.
Consequently, the reference amplitude entering Eq.~\eqref{eq:amplitude_ratio} becomes
\begin{equation}
\Psi_{\mathrm{ref}}(\bm{s})
=
\operatorname{sgn}(\tau_{\bm{s}})
\,\psi_{\mathrm{ref}}(\bm{s}).
\label{eq:mvmc_basis_sign}
\end{equation}
While this sign does not affect the sampling probability $\lvert\Psi_{\mathrm{ref}}(\bm{s})\rvert^2$, it must be accounted for in the phase of $R(\bm{s})$ in Eq.~\eqref{eq:amplitude_ratio}.

\section{Gradient-based Heisenberg circuit optimization}
\label{sec:direct_fidelity_optimization}

The tensor-network simulation also provides the gradients needed to optimize the circuit angles $\bm{\theta}$.
For each reference sample, we use boundary-MPS environments to obtain the derivatives of the circuit amplitude with respect to the site tensors $\bm{T}=\{T_i\}$.
Weighting and summing these derivatives according to Eq.~\eqref{eq:reference_fidelity_estimator} yields the tensor gradient $\nabla_{\bm{T}}F_{\mathrm{ref}}$.
Reverse-mode differentiation then propagates this gradient through the bond truncations and gate applications to obtain $\nabla_{\bm{\theta}}F_{\mathrm{ref}}$~\cite{Liao2019Differentiable}, while implicit differentiation accounts for how the converged BP messages $\bm{M}$ depend on the site tensors~\cite{Burgelman2026implicit}.

\begin{figure}[htbp]
\centering
\includegraphics[clip,width=1.0\linewidth]{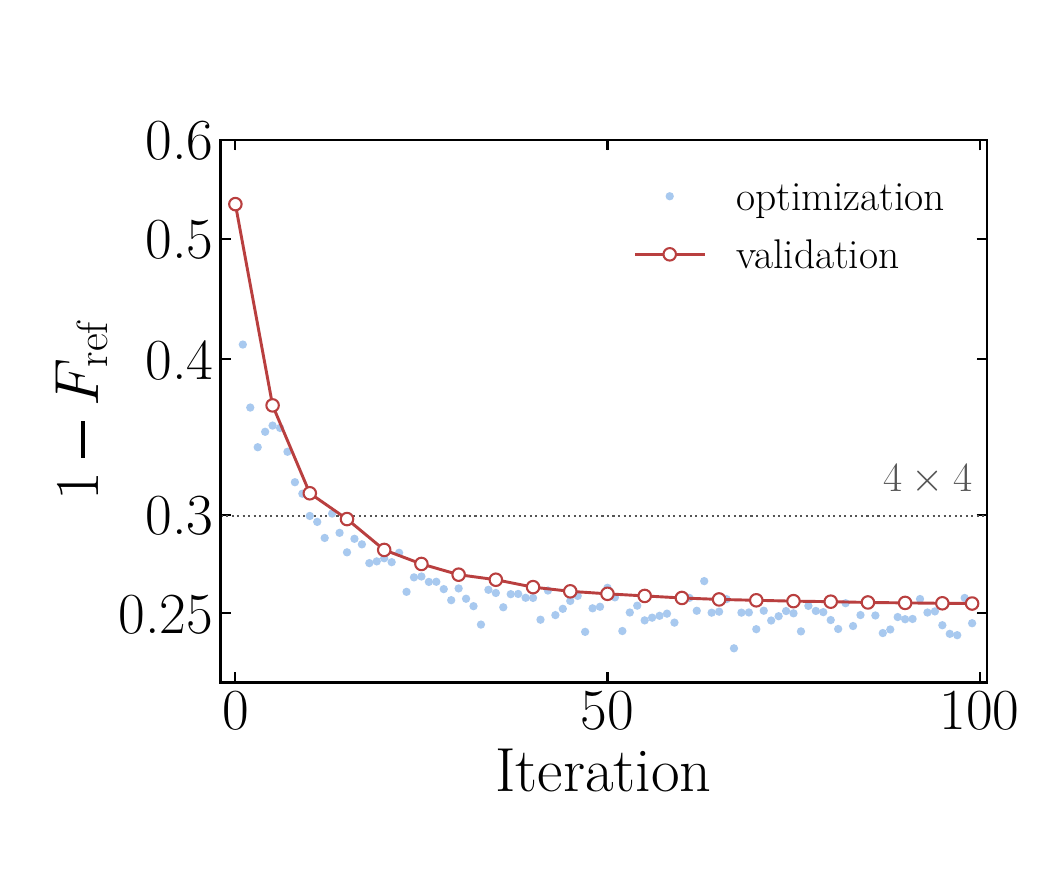}
\caption{Optimization of the $6\times6$ circuit with $N_{\mathrm b}=3$.
Blue and red points show $1-F_{\mathrm{ref}}$ for optimization and validation, respectively.
The dotted line marks the parameters transferred from the $4\times4$ lattice.}
\label{fig:fidelity_fit_l6_orbit_lanczos}
\end{figure}

\subsection{Benchmark on the \texorpdfstring{$6\times6$}{6x6} lattice}

As a further demonstration, we optimize the $L=6$, $N_{\mathrm b}=3$ circuit using the gradients described above.
The angles within each layer vary independently subject to lattice reflection symmetries, yielding 54 independent parameters.
We initialize these parameters with the layer-shared values perturbed by Gaussian noise with a standard deviation of $0.1$ rad, and optimize them with respect to the reference fidelity using the Adam optimizer, decaying the learning rate from $0.05$ to $0.001$ over 100 iterations.
We perform this optimization at $D=32$ and $\chi=64$, resampling 5,000 configurations from the VMC reference at each step.
For validation, we evaluate the fidelity at $\chi=96$ every five iterations as well as upon convergence, using a fixed set of 5,000 configurations from a separate reference sample.

Figure~\ref{fig:fidelity_fit_l6_orbit_lanczos} shows the resulting optimization trajectory.
To benchmark performance, we evaluate both the initial and optimized circuits at $D=48$ and $\chi=144$, matching the bond dimensions used in the parameter-transfer benchmark.
Through this optimization, the reference fidelity improves from $0.7055\pm0.0026$ with the 12 layer-shared angles to $0.7570\pm0.0024$ with the 54 optimized angles.
This result confirms that introducing additional independent angles and optimizing them on the target lattice enhances fidelity without requiring extra gates.

\end{document}